\documentclass[12pt]{article}

\usepackage{scicite}
\usepackage{graphicx}
\usepackage{times}
\usepackage{amsmath}
\usepackage{placeins}
\usepackage{xcolor}

\newenvironment{sciabstract}{%
\begin{quote} \bf}
{\end{quote}}

\usepackage[style=nature, doi=false,isbn=false,url=false,eprint=false,arxiv=false, backend=biber]{biblatex}
\AtEveryBibitem{\clearfield{month}}
\AtEveryBibitem{\clearfield{day}}
\AtEveryBibitem{\clearfield{note}}
\AtEveryBibitem{\clearlist{language}}
\bibliography{ms}

\title{An optoacoustic field-programmable perceptron for recurrent neural networks}

\author
{Steven Becker$^{1, 2}$, Dirk Englund$^{3}$, and Birgit Stiller$^{1, 2,\ast}$\\
\\
\normalsize{$^{1}$Max-Planck-Institute for the Science of Light,}\\
\normalsize{Staudtstr. 2, 91058 Erlangen, Germany}\\
\normalsize{$^{2}$Department of Physics, Friedrich-Alexander-Universität Erlangen-Nürnberg}\\
\normalsize{Staudtstr. 7, 91058 Erlangen, Germany}\\
\normalsize{$^{3}$Research Laboratory of Electronics, Massachusetts Institute of Technology,}\\
\normalsize{Cambridge, Massachusetts 02139, USA}\\
\\
\normalsize{$^\ast$ Corresponding author. Email:   birgit.stiller@mpl.mpg.de}
}

\date{}

\begin{document}


\baselineskip24pt


\maketitle


\begin{sciabstract}
A critical feature in signal processing is the ability to interpret correlations in time series signals, such as speech. Machine learning systems process this contextual information by tracking internal states in recurrent neural networks (RNNs), but these can cause memory and processor bottlenecks in applications from edge devices to data centers, motivating research into new analog inference architectures. But whereas photonic accelerators, in particular, have demonstrated big leaps in uni-directional feedforward deep neural network (DNN) inference, the bi-directional architecture of RNNs presents a unique challenge: the need for a short-term memory that (i) programmably transforms optical waveforms with phase coherence , (ii) minimizes added noise, and (iii) enables programmable readily scales to large neuron counts. Here, we address this challenge by introducing an optoacoustic recurrent operator (OREO) that simultaneously meets (i,ii,iii). Specifically, we experimentally demonstrate an OREO that contextualizes and computes the information carried by a sequence of optical pulses via acoustic waves. We show that the acoustic waves act as a link between the different optical pulses, capturing the optical information and using it to manipulate the subsequent operations. Our approach can be controlled completely optically on a pulse-by-pulse basis, offering simple reconfigurability for a use case-specific optimization. We use this feature to demonstrate a recurrent drop-out, which excludes optical input pulses from the recurrent operation.
We furthermore apply OREO as an acceptor to recognize up-to $27$ patterns in a sequence of optical pulses. Finally,
we introduce a DNN architecture that uses the OREO as bi-directional perceptrons to enable new classes of DNNs in coherent optical signal processing.
\end{sciabstract}

\FloatBarrier
\section*{Introduction}
Understanding the context of a situation is a powerful ability of the human brain, allowing it to predict possible outcomes and to make intelligent decisions. While humans can access the context of a situation via the short-term memory, machines struggle in contextualizing.
Artificial neural networks, one of the most powerful computing architectures,
face this problem as well. To overcome this limitation, they can be
equipped with recurrent feedback, allowing them to process current inputs
based on previous ones. The so-called recurrent neural networks (RNNs)
can contextualize, recognize, and predict sequences of information and
are applied for numerous applications such as language processing tasks, and for video and image processing~\autocite{yu_review_2019, salehinejad_recent_2018, van_den_oord_pixel_2016, mesnil_using_2015, donahue_long-term_2017}.
One of the simplest versions of a RNN is the Elman network~\autocite{elman_finding_1990}, which
adds a recurrent operation to each neuron of its fully-connected network, analogous to the neuron's activation function.
With this three-layer network,~\citeauthor{elman_finding_1990} was already able to understand simple grammatical structure.
More complex models have proven themselves as Chinese poets, rap artists, and empathetic listeners~\autocite{zhang_chinese_2014, potash_ghostwriter_2015, lee_high-level_2015}.

Currently, the scientific community aims to transfer electronic neural networks into the optical domain.
The resulting optical neural networks have attracted great interest due
to their promises of high processing speed and broad bandwidth, and low dissipative losses~\autocite{shastri_photonics_2021, bogaerts_programmable_2020, wetzstein_inference_2020}.
Thus, they are considered to pave the way towards energy efficient and highly
parallel optical circuits, enhancing the performance and
capabilities of future artificial neural networks~\autocite{shen_deep_2017, tegin_scalable_2021, zuo_all-optical_2019, lin_all-optical_2018, feldmann_all-optical_2019, zhang_optical_2021, chen_deep_2023}.

Although the field of optical neural networks has made great progress in recent years, the field of recurrent optical neural networks is still very limited to concepts based on artificial reservoirs, such as free-space
cavities~\autocite{bueno_reinforcement_2018}, delay systems~\autocite{brunner_tutorial_2018, mourgias-alexandris_all-optical_2020}, and microring resonators~\autocite{tait_neuromorphic_2017}.
These designs can face several challenging issues.
Firstly, the usage of an artificial cavity, e.g. a ring resonator, can limit the scalability of those networks.
Secondly, the cavity may not be frequency sensitive, preventing them from being applied for resource-efficient multi-frequency data processing.
Finally, the cavity's recurrent weights cannot be varied rapidly, limiting the control of the recurrent process such as the implementation of recurrent dropout on single pulse level in order to regularize the network.

\begin{figure}[h!]
\centering
  \includegraphics[width=\textwidth]{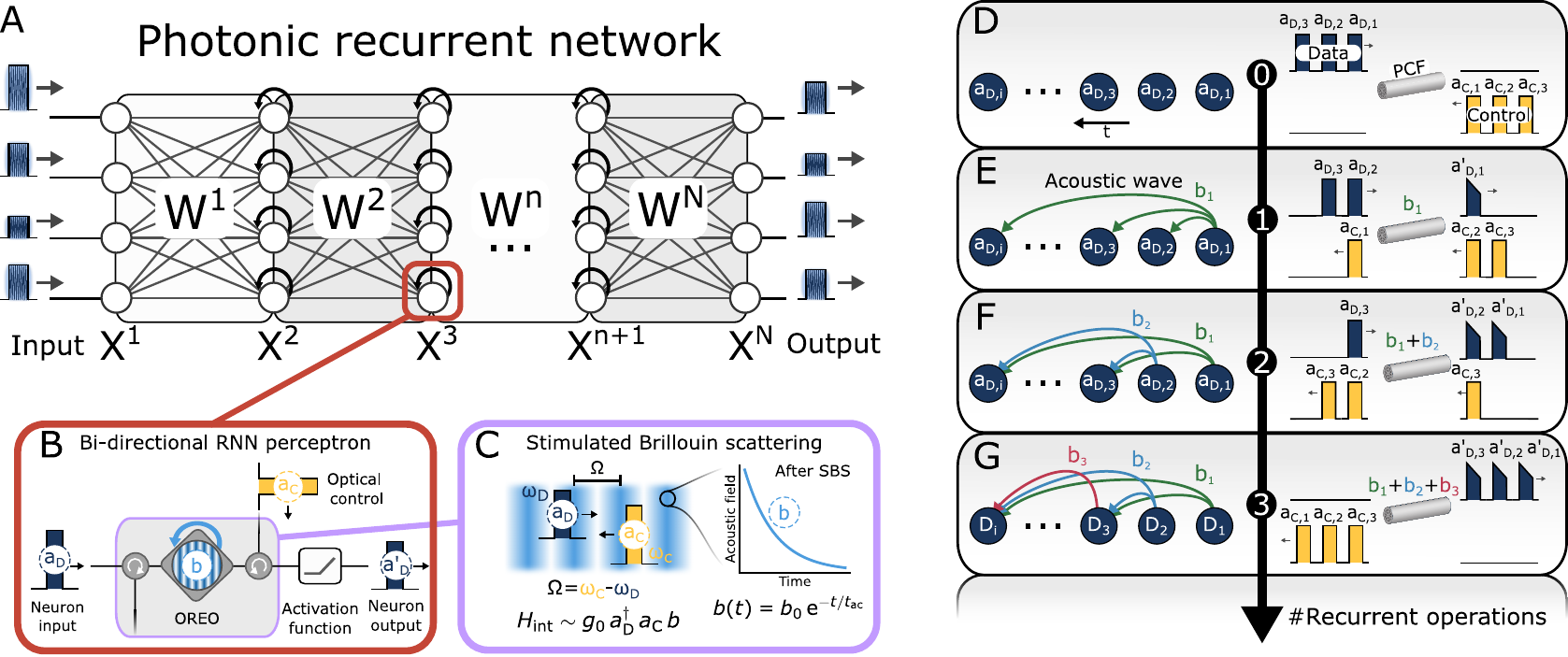}
    \caption{\color{black} Schematic of the optoacoustic recurrent operator (OREO) and its proposed function in a recurrent NN.
    \textbf{A} - An example of a photonic recurrent network with $N$ layers $\mathrm{X^n}$, which are connected by a matrix operation $\mathrm{W^n}$. The blue arrow indicates the recurrent nature of one neuron.
    \textbf{B} - The bi-directional perceptron contains an OREO and an activation function. OREO captures and links sequential information $a_\mathrm{D}$ using a sound wave $b$, which is generated by SBS and controlled by an optical control pulse $a_\mathrm{C}$. The output of the acoustic recurrent neuron $a'_\mathrm{D}$ is fed into the next layer of the neural network.
    \color{black}
    \textbf{C} - Conceptional illustration of the SBS process with its interaction Hamiltonian $H_\mathrm{int}$. The sound wave $b$ carries the information of the neuron's input and decays after the SBS process.
    \textbf{D} - \textbf{G}: Illustration of three recurrent operations performed by OREO.
    \textbf{D} shows the initial situation with three data-control pulse pairs separated by a deadtime $dt$.
     The data and control pulses are launched from opposite sides into a photonic crystal fiber (PCF).
     \textbf{E} shows the system after the SBS-interaction of $a_\mathrm{D,1}$ and $a_\mathrm{C,1}$, which transfers energy from $a_\mathrm{D,1}$
     to an acoustic wave $b_1$. \textbf{F} shows the system after a second pulse pair has passed the PCF.
     The acoustic wave $b_1$ connects the interaction of $a_\mathrm{D,2}$ and $a_\mathrm{C,2}$ with the previous one, while the SBS process transfers
     information from $a_\mathrm{D, 2}$ into $b_2$. \textbf{G} highlights the acoustic link created by OREO between three optical pulse pairs.
    }
  \label{fig: illustration_introduction}
\end{figure}

%
Here, we experimentally demonstrate an optoacoustic recurrent operator (OREO) based
on stimulated Brillouin-Mandelstam scattering (SBS) that can unlock recurrent functionalities in existing optical neural network architectures~(see Fig.~\ref{fig: illustration_introduction} \textbf{A}).
SBS is an interaction of optical waves with traveling acoustic waves which serve in our system as a latency component due to the slow acoustic velocity.
OREO is therefore able to contextualize a time-encoded stream of information by using acoustic waves as a memory to remember previous operations~(see Fig.~\ref{fig: illustration_introduction}\textbf{B}).

In contrast to previously reported approaches \autocite{brunner_tutorial_2018, bueno_reinforcement_2018, hughes_wave_2019, mourgias-alexandris_all-optical_2020, tait_neuromorphic_2017},
OREO controls its coherent recurrent operation completely optically on pulse level
without the need of any artificial reservoir such as a ring resonator or a delay system.
Hence, OREO does not rely on complicated manufacturing processes of microstructures.
It functions in any optical waveguide, including on-chip devices, as it harvests
the physical property of a sound wave~\autocite{zhu_stored_2007, merklein_chip-integrated_2017, stiller_-chip_2018}. In particular, with the announcement of
the first on-chip EDFA~\autocite{liu_photonic_2022} a fully integrated design
is seemingly close.
%

%
%

%
We demonstrate OREO experimentally from different perspectives.
Firstly, we show how OREO links different input states of subsequent optical pulses
to each other via acoustic waves. Secondly, we present how the all-optical control of OREO can be used to implement a recurrent dropout.
Finally, we apply OREO as an acceptor~\autocite{goldberg_neural_2022} to predict up to $27$ different patterns carried by
a time series of input pulses.

%
%

\FloatBarrier
\section*{Concept of an optoacoustic recurrent operator}
The recurrent operation of OREO is based on the interaction of optical and acoustic waves through SBS, which is one of the most prominent third-order nonlinear effects and describes
the coherent coupling of two optical waves, data and control, to an acoustic wave in a material. The dynamic is illustrated in Figure~\ref{fig: illustration_introduction}\textbf{C} and follows from the Hamiltonian~\eqref{eq: SBS_Hamilton}~\autocite{j_e_sipe_hamiltonian_2016, zhang_quantum_2023, eggleton_brillouin_2022}:
\begin{align}
    \label{eq: SBS_Hamilton}
\begin{aligned}
    H_\mathrm{OREO} = \, &\hbar \omega_\mathrm{D} \int_{-\infty}^{\infty} \mathrm{d}z \, a^{\dagger}_\mathrm{D}(z,t)a_\mathrm{D}(z,t) +
                                  \hbar \omega_\mathrm{C} \int_{-\infty}^{\infty} \mathrm{d}z \, a^{\dagger}_\mathrm{C}(z,t)a_\mathrm{C}(z,t) \, +\\
                      +\,&\hbar \Omega  \int_{-\infty}^{\infty} \mathrm{d}z \, b^{\dagger}(z,t)b(z,t) +
                      \underbrace{
                       \hbar g_0 \, \int_{-\infty}^{\infty} \mathrm{d}z \, \left(a_\mathrm{D}(z,t)\, a^{\dagger}_\mathrm{C}(z, t) \, b^{\dagger}(z, t) + \, \mathrm{H.c.}\right)}_{\mathrm{Interaction \, \, Hamiltonian}},
\end{aligned}
\end{align}
using the optoacoustic coupling constant $g_0$, the frequency relation between the optical fields $\omega_\mathrm{D} = \omega_\mathrm{C} + \Omega$, and the wave packet operators $a_\mathrm{D}$, $a_\mathrm{C}$, $b$ of the data, control and acoustic field, respectively.
Similar to the clinking of a wine glass, the acoustic wave $b$ persists beyond its excitation,
decaying exponentially with time
$b(t)\propto \exp\left(-t/\tau_\mathrm{ac}\right)$, where $\tau_\mathrm{ac}\propto\Gamma_\mathrm{ac}^{-1}$ is the acoustic lifetime, which depends on the properties of the used waveguide and is for a photonic crystal fiber (PCF) about $\tau_\mathrm{ac}\approx 10\,\mathrm{ns}$ (see Fig.~\ref{fig: illustration_introduction}\textbf{C}).
As a result, an acoustic wave $b_i$ can seed subsequent SBS processes $j>i$. Moreover, the acoustic builds up with each SBS process, which can be described as a superposition of all previous created acoustic waves $b_i$ with amplitude $b_{0, i}$, created at the time $t_i$ and carrying a phase $\varphi_i$. Hence, the acoustic wave $b_N$ after $N$ SBS interactions:

\begin{equation}
    \label{eq: total_acoustic_field}
    b_N(z, t) =  \sum_{i=1}^{N} b_i(z, t) = \sum_{i=1}^{N} b_{0, i}(z)\,  e^{-\frac{t-t_i}{\tau_\mathrm{ac}}+\mathrm{i}\varphi_i},
\end{equation}
yields the recurrence in the interaction Hamiltonian:
\begin{align}
    \label{eq: recurrent_interaction_Hamilton}
    \begin{aligned}
    H_\mathrm{int, N} = \hbar g_0 \, \int_{-\infty}^{\infty} \mathrm{d}z \, \left(a_\mathrm{D}(z,t)\, a^{\dagger}_\mathrm{C}(z, t) \, \left( \sum_{i=1}^{N} b^{\dagger}_i(z, t) \right) + \, \mathrm{H.c.}\right)
    \end{aligned}
\end{align}
\color{black}
Equation~\eqref{eq: recurrent_interaction_Hamilton} shows furthermore that programming the field $a_\mathrm{C}$ controls the acoustic feedback all-optically, enabling
a pulse-by-pulse increase or suppression.
For instance, setting $a_\mathrm{C, i}=0$ corresponds to a recurrent dropout so that $a_\mathrm{D, i}$ leaves the fiber unchanged.
\color{black}

\color{black}
We experimentally implement OREO in a telecom-fiber apparatus illustrated in Figure~\ref{fig: illustration_introduction} \textbf{D}.
Here we launch several consecutive optical input data pulses $a_\mathrm{D, i}$
and strong counter-propagating optical control pulses $a_\mathrm{C, i}$ into a PCF.
\color{black}
The optical data pulses are shifted up in frequency by $\Omega\approx 10.6\,\mathrm{GHz}$
compared to the optical control pulses, which is close to the Brillouin frequency of the PCF.
When a data and control pulse pair $a_\mathrm{D, 1}$ and $a_\mathrm{C, 1}$ meets inside the PCF, they induce SBS,
depleting the data pulse and transferring its energy into the acoustic domain. Eventually, an acoustic wave $b_1$ is generated, which persists much longer than the optical interaction (see Figure~\ref{fig: illustration_introduction} \textbf{E}).
An optoacoustic recurrent operation is performed, when a subsequent data and control pulse pair ($a_\mathrm{D, 2}$ and $a_\mathrm{C, 2}$)
reaches the acoustic wave $b_1$ before it has decayed. Hence, the deadtime $dt$ until the second pulse pair arrives
must be less then the acoustic lifetime. The previously generated acoustic wave
connects to the subsequent SBS process between $a_\mathrm{D, 2}$ and $a_\mathrm{C, 2}$ and establishes a link of the second data pulse $a_\mathrm{D, 2}$ to the first data pulse $a_\mathrm{D, 1}$. In addition, the second SBS process creates a second acoustic wave $b_2$, carrying information of $a_\mathrm{D, 2}$.
Now, the acoustic domain holds information of both data pulses $a_\mathrm{D, 1}$ and $a_\mathrm{D, 2}$ (see Figure \ref{fig: illustration_introduction} \textbf{F}).
The discussed procedure could now be repeated also for a third pulse pair and, in general, as long as
subsequent pules pairs arrive before the acoustic wave decayed completely (see Figure \ref{fig: illustration_introduction} \textbf{G}).
\color{black}
%
\section*{Experimental results}
\begin{figure}[h!]
\centering
  \includegraphics[width=.9\textwidth]{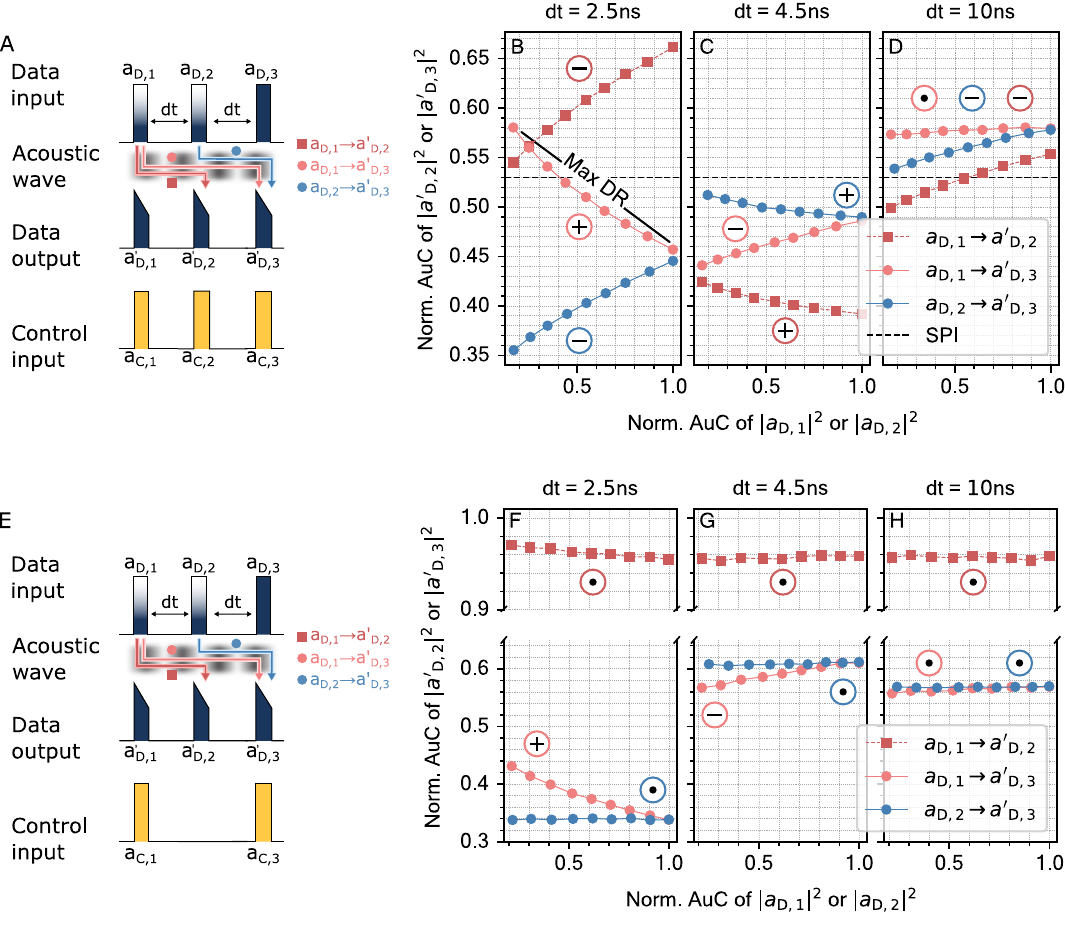}
    \caption{ Observing OREO's optoacoustic linking and recurrent dropout capabilities. \textbf{A} - Schematic illustration of the amplitude sweep that investigates how different optical states are passed between the optical data pulses $a_\mathrm{D, i}$ via an acoustic wave $b$.
    \textbf{B} to \textbf{D} - Experimental results of the amplitude sweep. While $a_\mathrm{D, i}, \, i=1,2$ is changed, its impact on the subsequent pulses $a_\mathrm{D, j}, \, j>i$ is studied for different deadtimes $dt$. Each SBS process creates an acoustic wave $b_i$, which interferes with pre-exisiting ones $b_k, \, k<i$, eventually. We mark the links $a_\mathrm{D, i} \to a_\mathrm{D, j}$ with a $+$, $-$, and $\cdot$, when they experienced an enhancement, a reduction and an annihilation of the SBS process, respectively. We added the depletion of a single pulse interaction (SPI) as reference.
    \textbf{E} - Schematic illustration of the pulse configuration used to study the OREO's feature to implement a recurrent dropout.
    \textbf{F} to \textbf{H} - Experimental results of OREO's recurrent dropout capabilities.
    }
    \label{fig: results_amp_sweep_pulse_skipping}
\end{figure}

In the following, we study the acoustic link by sweeping the data pulse amplitude of either $a_\mathrm{D, 1}$ or $a_\mathrm{D, 2}$, while keeping the corresponding subsequent pulses constant.
For instance, if the input amplitude of $a_\mathrm{D, 1}$ is varied, $a_\mathrm{D, 2}$ and $a_\mathrm{D, 3}$ are fixed in amplitude.
The control pulses $a_\mathrm{C, i}$ are kept constant over the entire study. For each amplitude step, we measure the pulse's area under curve (AuC) of the output pulses $a'_\mathrm{D, i}$.
An AuC-measurement of $a_\mathrm{D, i}$ without control pulses serves as reference. In total, three different acoustic links
occur from this experimental configuration, namely, $a_\mathrm{D, 1} \to a'_\mathrm{D, 2}$, $a_\mathrm{D, 1} \to a'_\mathrm{D, 3}$, and $a_\mathrm{D, 2} \to a'_\mathrm{D, 3}$ (see Figure~\ref{fig: results_amp_sweep_pulse_skipping} \textbf{A}). In order to rule out drifting effects, we measure each amplitude twice in a random order and take the mean value afterwards.
Furthermore, the amplitude sweep is performed for three different time delays $dt=2.5,\, 4.5, \, \mathrm{and}\, 10\,\mathrm{ns}$, as
the acoustic link decays over time.
For a deadtime of $2.5 \, \mathrm{ns}$, an increase in amplitude of $a_\mathrm{D, 1}$ raises the output amplitude $a'_\mathrm{D, 2}$ as shown by Figure~\ref{fig: results_amp_sweep_pulse_skipping} \textbf{B}.
Because the degree of depletion is lower as for a single pulse interaction (SPI), e.g., $a_\mathrm{D, 1}\to a'_\mathrm{D, 3}$, we conclude that the acoustic wave $b_1$
weakens the SBS process of $a_\mathrm{D, 2}\leftrightarrow a_\mathrm{C, 2}$.
This finding can be explained with the different acoustic phases (see equation~\eqref{eq: total_acoustic_field}), which can lead to constructive or destructive interference of the acoustic waves during the SBS process, The acoustic phase is introduced by detuning the frequency difference between data and control pulses slightly from the Brillouin frequency.
The acoustic interference is also the reason for the decreasing behavior of the link  $a_\mathrm{D, 1} \to a'_\mathrm{D, 3}$, here, the acoustic wave $b_1$ enhances the
SBS process of $a_\mathrm{D, 2}\leftrightarrow a_\mathrm{C, 2}$. The symbols $\mathbf{+}$ and $\mathbf{-}$ mark the constructive and destructive nature of the underlying acoustic interference in Figure~\ref{fig: results_amp_sweep_pulse_skipping} \textbf{B}, respectively.
OREO achieves a maximal dynamic range (Max DR) of $33\,\%$.
For a deadtime of $dt = 4.5 \, \mathrm{ns}$ we observe a flip in the dynamic as all links switch their behavior from a constructive ($\mathbf{+}$) to destructive ($\mathbf{-}$) acoustic link, and vice versa (depicted in Figure~\ref{fig: results_amp_sweep_pulse_skipping} \textbf{C}). In addition, the overall level of depletion is larger in comparison to the SPI-case.
For a deadtime of $dt = 10 \, \mathrm{ns}$ (equal to the acoustic lifetime), the dynamic range of the optical connection decreases further as we can see for the connection $a_\mathrm{D, 1} \to a'_\mathrm{D, 2}$ in Figure~\ref{fig: results_amp_sweep_pulse_skipping} \textbf{D}. Ultimately, the effect of the decaying acoustic wave becomes in particular visible for the interaction $a_\mathrm{D, 1} \to a'_\mathrm{D, 3}$ as $a'_\mathrm{D, 3}$ remains constant over the entire sweep range of $a_\mathrm{D, 1}$ (see Fig~\ref{fig: results_amp_sweep_pulse_skipping} \textbf{D}). Note that we marked vanished acoustic links
with the $\bullet$-symbol.

With this, we have shown that OREO connects the information carried by subsequent optical data pulses.
The acoustic link is sensitive to the amplitude and deadtime of the involved optical data pulses. As
the interaction is continuous, it can be used for digital and analogue recurrent tasks.
Moreover, the acoustic interference observed with OREO ties in with previous studies based on continuous optical waves and our measurement extends the observation of acoustic interference into a pulsed context~\autocite{feng_coherent_2019, okawa_optical_2020}.
In the supplementary material, we study OREO numerically and experimentally in a highly nonlinear fiber (HNLF),
using the framework presented in Reference~\autocite{de_sterke_nonlinear_1991}.
With the HNLF we study the linear response of OREO, which occurs in the case
that the frequency difference of data and control matches exactly the Brillouin frequency.

%
%
%
OREO controls the recurrent operation completely optically via the control pulses, enabling us to implement use case specific computations.
For instance, in a pulse sequence consisting of three data pulses, one could
skip the middle pulse $a_\mathrm{D, 2}$ by dropping the second control pulse, which could be useful for regularization~\autocite{semeniuta_recurrent_2016}.
In order to demonstrate the recurrent dropout, we excluded the second control pulse $a_\mathrm{C, 2}$ from the pulse train.
Note, that the amplitudes of the other control pulses remain the same. In a next step, we vary
the amplitude of data pulses $a_\mathrm{D, 1}$ and $a_\mathrm{D, 2}$ in upward and downward direction and check the impact
on the subsequent data pulses
(see Figure~\ref{fig: results_amp_sweep_pulse_skipping} \textbf{E}).
Furthermore, we change the deadtime to investigate the influence of the acoustic interference
on the interaction $a_\mathrm{D, 1}\to a'_\mathrm{D, 3}$.

OREO turns off the links between $a_\mathrm{D, 1} \to a'_\mathrm{D, 2}$ and  $a_\mathrm{D, 2} \to a'_\mathrm{D, 3}$ as we can see in
in Figure~\ref{fig: results_amp_sweep_pulse_skipping} \textbf{F}. As marked with the
$\bullet$-symbol, those two links show a constant behavior for the entire amplitude sweep.
Only the interaction $a_\mathrm{D, 1}\to a'_\mathrm{D, 3}$ is active as the control pulses $a_\mathrm{C, 1}$ and $a_\mathrm{C, 3}$ establish the required acoustic
link. Note, that for the case of $a_\mathrm{D, 2} \to a'_\mathrm{D, 3}$ the interaction $a_\mathrm{D, 3} \leftrightarrow a_\mathrm{C, 3}$ is influenced
by the acoustic wave generated of the $a_\mathrm{D, 1} \leftrightarrow a_\mathrm{C, 1}$-interaction ($a_\mathrm{D, 1}$ is constant).
This link can also explain the lower degree of depletion of $a_\mathrm{D, 1}\to a'_\mathrm{D, 3}$ at $dt=4.5 \, \mathrm{ns}$ (see Figure~\ref{fig: results_amp_sweep_pulse_skipping}~\textbf{G}).
Here, the $a_\mathrm{D, 1}$ and $a_\mathrm{D, 3}$ are already separated by $10 \, \mathrm{ns}$, which eliminated almost their acoustic link.
At a deadtime of $dt = 10 \,\mathrm{ns}$, the $a'_\mathrm{D, 3}$ is completely disconnected from $a_\mathrm{D, 1}$ and $a_\mathrm{D, 2}$ as can be seen
by the constant behavior of $a'_\mathrm{D, 3}$ for both sweeps of $a_\mathrm{D, 1}$ and $a_\mathrm{D, 2}$ (see Figure~\ref{fig: results_amp_sweep_pulse_skipping}~\textbf{H}).
Besides, over all measurements, $a'_\mathrm{D, 2}$ is below the reference level ($a'_\mathrm{D, 2} < 1$), e.g., for the interaction $a_\mathrm{D, 1} \to a'_\mathrm{D, 2}$ in Figure~\ref{fig: results_amp_sweep_pulse_skipping} \textbf{F}.
The increased optical noise floor appears as soon as the EDFA is turned on and could lead to this intrinsic depletion.
%
%
%
%
%
%
\FloatBarrier
\subsection*{Optical pattern recognition}
\begin{figure}[h!]
\centering
  \includegraphics[width=.8\textwidth]{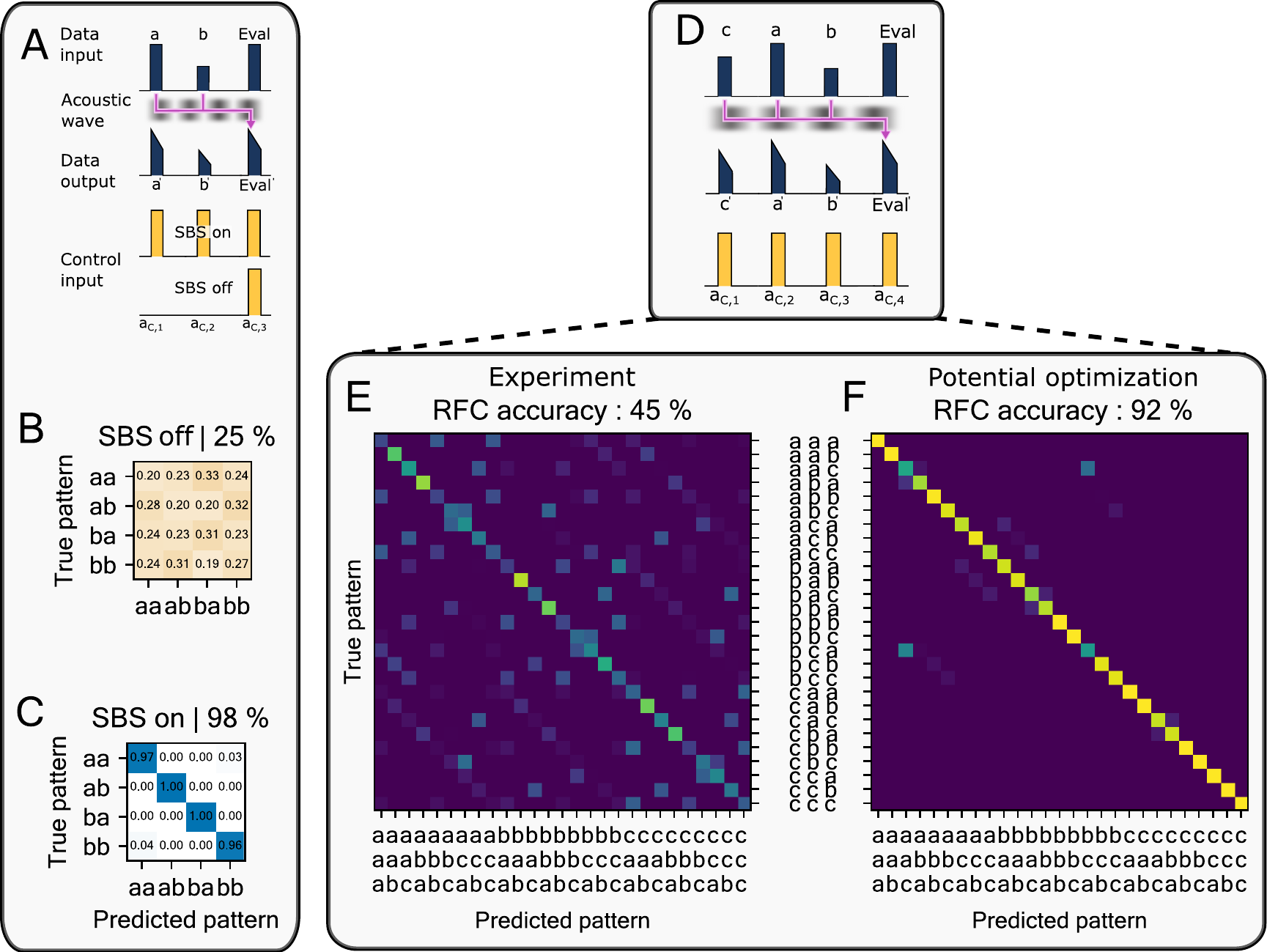}
    \caption{ Applying OREO as an acceptor to predict patterns of optical pulses.
    \textbf{A} - Schematic illustration of how the acoustic link can be used by an optical evaluation pulse (Eval) to predict a pattern from optical pulses, which have been launched into the optical fiber before. The example shows a $ab$-pattern. \textbf{B} and \textbf{C} - Confusion matrix of a \textsc{Random Forest} (RFC) which is used to classify the data set with and without SBS. The RFC achieves an almost perfect classification rate as soon as OREO provides the recurrent feedback.
    \textbf{D} - Schematic illustration of the three pulse pattern recognition task. This case shows the $cab$-pattern. \textbf{E} shows the confusion matrix of the RFC, using $30\,\%$ of the data set for training. The RFC achieves an accuracy of $45\, \%$ and outperforms a simple guess by $11$-times. The accuracy of the RFC is mostly limited by experimental precision.
    \textbf{F} - RFC confusion matrix using simulated data to study OREO's performance with experimental optimization. In this case, we can achieve an accuracy of $92\,\%$. The simulation are based on a frequency matched SBS process and is described in more detail in the supplementary material.
    } 
  \label{fig: pattern_recoginition}
\end{figure}

From the beginning on, recurrent operators have been used to recognize patterns~\autocite{elman_finding_1990}.
In the following section, we employ OREO as an acceptor~\autocite{goldberg_neural_2022} to recognize any
pattern that can be created with two different data pulses $a$ and $b$: $aa$, $ab$, $ba$ \& $bb$, where
the $b$-pulse is half the amplitude of the $a$-pulse.
Each pulse is launched with a matching control pulse $a_\mathrm{C, i}$ into the PCF, where SBS is introduced. The deadtime between
two consecutive pulses is $2.5\,\mathrm{ns}$. As a measure of OREO's performance, we launch
an evaluation pulse pair into the optical fiber and use the AuC of an output evaluation pulse (Eval)
(see Figure~\ref{fig: pattern_recoginition} \textbf{A}).
In total, we check all patterns $250$-times in a random order and classify the resulting
data set ($70\,\% \, \mathrm{training}$, $30\,\% \mathrm{testing}$) with a \textsc{Random Forest} classifier \autocite{tin_kam_ho_random_1995} (RFC) implemented in
\textsc{sklearn}-package~\autocite[v1.1.3]{pedregosa_scikit-learn_2011}. Furthermore, we perform the 
described study twice, once with the SBS-process and once without to isolate
OREO's effect.
When OREO is off, the RFC cannot distinguish the different patterns and shows the same accuracy as a random guess (see Figures~\ref{fig: pattern_recoginition} \textbf{B}). However with OREO, the RFC is capable of distinguishing the different patterns almost with an accuracy of almost a $100\,\%$ (see Figures~\ref{fig: pattern_recoginition} \textbf{C}).

Next, pushing OREO to the acoustic lifetime limit,
we evaluate its performance for three different states encoded
onto three pulses. The third state $c$ is three quarters of the $a$ state. In total,
we test OREO to distinguish every possible permutation of $a$, $b$, and $c$, giving
$27$ different patterns. This time we launch a fourth data-control pair into the PCF,
in order to evaluate OREO's memory (see Figure~\ref{fig: pattern_recoginition} \textbf{D}).
Note that all four control pulses $a_\mathrm{C, i}$ carry the same
optical energy as in the three pulse configuration. We increased the sample size $n$ per
pattern from $250$ to $500$ in order to decrease statistical errors.
Figure \ref{fig: pattern_recoginition} \textbf{E} shows the corresponding confusion matrix.
OREO functions as  acceptor and generates distinguishable distributions for the
$27$ patterns. The \textsc{RFC} achieves an accuracy of $45\,\%$, exceeding the accuracy of a simple guess by $11$-times.
Note that the classification task of the $abc$-case has $729$ possible classification outcomes, and is
$45$ times more complex as the $ab$-case. In general, it is seven times more complex as an image classification task
based on the \textsc{MNIST} dataset~\autocite{lecun_mnist_2010} with $10\times 10$ degrees.
The performance of OREO is currently limited by experimental precision,
which is reduced by drifts of the optical pulses over the measurement period.
Therefore, we perform a numerical analysis of OREO as an acceptor in the frequency matched case, in order to assess its potential performance. In this simulated experiment, OREO and the RFC achieve an accuracy
of $92 \,  \%$. Figure~\ref{fig: pattern_recoginition} \textbf{F} shows the corresponding confusion matrix.
In the supplementary material, we describe the numerical analysis and check the impact of the pulse width, deadtime, acoustic lifetime, and experimental precision on OREO's pattern recognition performance. This analysis indicates that OREOs performance can even be pushed further to an accuracy of $97\,\%$.
\FloatBarrier
\section*{Discussion and future possibilities}
The acoustic link employed by
OREO enables the processing of time-encoded serial information within a PCF. Its capability to control the
recurrent interaction all-optically, gives the concept unique features. The adjustable amplitudes of the control
pulses allow OREO's behavior to be changed at the single pulse level, offering an all-optical degree of freedom to adjust
its recurrent operation. Moreover, we have shown that it offers the possibility to exclude data pulses from the
recurrent interaction. As a consequence, a single data pulse can propagate through OREO without experiencing any manipulation.
This can be used to implement recurrent dropout as regularization for the RNN.

The coherent nature of the underlying SBS process offers OREO not only to compute amplitude information
but also phase information. Eventually, OREO could compute quadrature amplitude modulated (QAM) data streams.
Higher memory depths could be achieved with three different approaches.
Firstly, a higher pulse density could be used to increase the number of operations that could be performed
within the intrinsic acoustic lifetime. This could be achieved by decreasing the
pulse width and the deadtime between the pulse pairs. For instance, with a
pulse width of $100\,\mathrm{ps}$ and a deadtime of $100\,\mathrm{ps}$ (the minimal
deadtime is dictated by the length of the waveguide), one could induce up to $50$ recurrent
interactions. Secondly, one could increase the acoustic lifetime to realize a
deep recurrent link, for instance by using materials with longer acoustic lifetimes or operating at cryogenic temperatures.
Thirdly, an optical refreshment of the acoustic waves could lead to an increase in memory depth~\autocite{stiller_coherently_2020}.
Because the SBS process does not significantly change the optical control pulses,
an optical recycling scheme could be applied to achieve high computational efficiencies.
\color{black}
Computational efficiency is determined by the number of operations (OPS) that OREO can perform with
one Joule of power. With an optical recycling scheme this value depends only on the deadtime
between the pulse pairs, yielding an efficiency from up to $\approx 11 \frac{\mathrm{POPS}}{\mathrm{J}}$; it could potentially increase  the computational efficiency of the method described in Reference~\autocite{mourgias-alexandris_all-optical_2020} by three orders of magnitude.
\color{black}
A more detailed description of the computational efficiency can be found in
the supplementary material.
The information bandwidth of an optical signal can be significantly increased by employing different optical frequencies as independent information channels.
This has been recently exploited by \citeauthor{sludds_delocalized_2022}~\autocite{sludds_delocalized_2022} to implement an
high-performance optical deep learning architecture for edge computing. OREO could be added to this scheme as SBS
is highly frequency-selective~\autocite{stiller_cross_2019}.
This unique feature of the optoacoustic interaction could also
be employed together with an optical multi-frequency matrix
operator~\autocite{davis_iii_frequency-encoded_2022, feldmann_parallel_2021, buddhiraju_arbitrary_2021} to realize an
multi-frequency recurrent neural network.

\section*{Conclusion}
In conclusion, we have demonstrated the first optoacoustic recurrent operator (OREO), which
connects the information carried by subsequent optical data pulses.
Our work combines for the first time the field of traveling acoustic waves and artificial neural networks and paves the way towards SBS-enhanced computing platforms. This new fusion brings context to optical neural networks, but can also enable much more. Typical building-blocks of a neural network, such as
nonlinear activation functions and other types of optoacoustic operators are within reach. Especially, the different time scales of optical and acoustic waves open up a whole new playground for the implementation of a variety of computing architectures.


\section*{Methods}
\begin{figure}[h!]
\centering
  \includegraphics[width=1\textwidth]{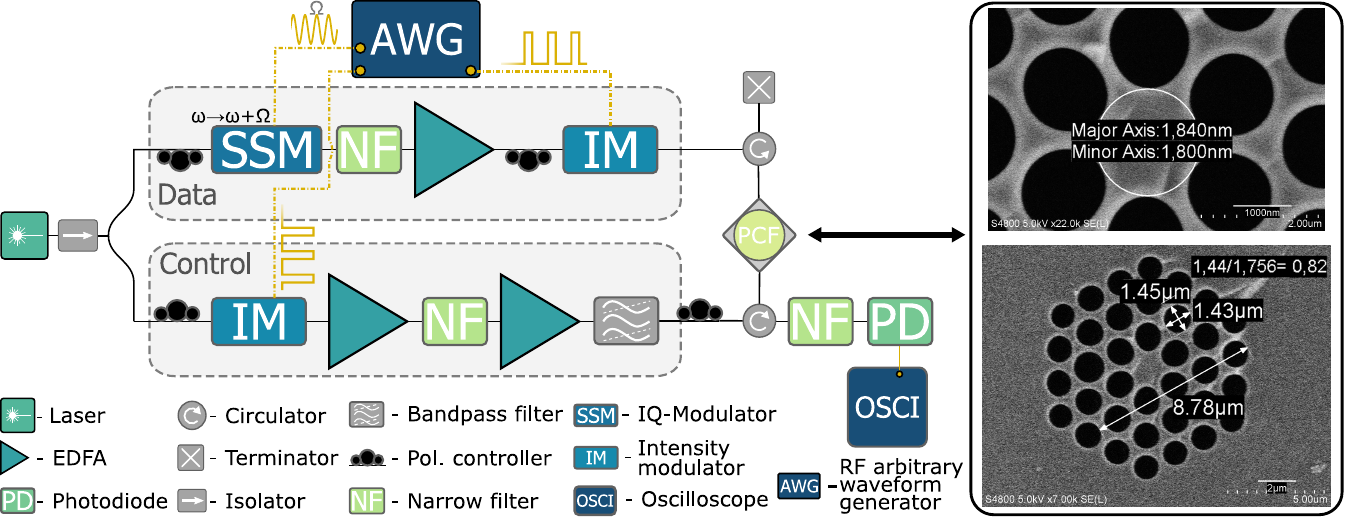}
    \caption{Illustration of the setup used to demonstrate the recurrent optoacoustic operator.
            The bandwidth of the oscilloscope and photodiode are $16\,\mathrm{GHz}$ and $12\,\mathrm{GHz}$, respectively. We introduce the Brillouin process inside  photonic crystal fiber (PCF), which could be replaced with any other waveguide or an on-chip device.
            }
\label{fig: illustration_setup}
\end{figure}

To demonstrate OREO, we build the all-fiber setup shown in Figure~\ref{fig: illustration_setup}.
As a sample, we use a photonic crystal fiber (PCF) with a length of $\approx 40 \,\mathrm{cm}$, an average hole diameter of $1.44\,\mu\mathrm{m}$, an average core diameter of $1.842\,\mathrm{nm}$, a pitch of $1.756\,\mu\mathrm{m}$, and $d/\Lambda=0.82$.
A continuous wave laser at $1550\,\mathrm{nm}$ is split into the data and control branch via a 50/50-splitter.
An IQ-modulator shifts the data signal by $\Omega\approx 10.6\,\mathrm{GHz}$, which is close to the PCF's Brillouin frequency of $\Omega_\mathrm{PCF}\approx 10.45\,\mathrm{GHz}$.
The data signal's spectrum is cleaned with
a subsequent narrow bandpass filter and afterwards amplified by an Erbium-doped fiber amplifier (EDFA).
An optical intensity modulator driven by an arbitrary waveform generator (AWG)
generates the optical pulses and, thus,
imprints the amplitude-encoded information. A single data pulse is $1\,\mathrm{ns}$
long and separated to an adjacent data pulse by a deadtime $dt$. The repetition
rate of a pulse sequence is $\approx 1\,\mathrm{MHz}$. The pulses are guided to the
PCF by an optical circulator and, afterwards, measured with a high-speed photodiode and
a $16\,\mathrm{GHz}$ Oscilloscope. The optical power of the data pulse is about
$1\,\mathrm{mW}$. An additional narrow bandpass filter cleans the signal before detection.
In the control branch, optical pulses are generated with the same pulse width and
repetition rate as the data branch.
Afterwards, the pulsed signal is amplified by an EDFA and filtered by a
narrow bandpass filter before launched into a high-power EDFA. The amplified signal is
filtered by a $1\,\mathrm{nm}$-width bandpass filter and launched with an average power of about $126\,\mathrm{mW}$
into the SBS process.

\section*{Acknowledgments}
The authors thank Florian Marquardt, Christian Wolff, \color{black}Changlong Zhu\color{black}, and Jesús Humberto Marines Cabello for fruitful discussions.
\textbf{Funding:} We acknowledge funding from the Max Planck Society through the Independent Max Planck Research Groups scheme and the Studienstiftung des deutschen Volkes.
\newline
\textbf{Author contributions:} Conceptualization: S.B. and B.S. Methodology: S.B., D.E., and B.S. Investigation: S.B. Visualization: S.B., D.E., and B.S.  Funding acquisition: B.S. Project administration: B.S. Supervision: B.S. Writing: S.B., D.E., and B.S.
\newline
\textbf{Competing interests:} S.B. and B.S. have filed a patent related to OREO: EP23153328.2.

\nocite{boyd_nonlinear_2008, agrawal_nonlinear_2013, eggleton_brillouin_2022, eggleton_brillouin_2022-1}
\nocite{buckland_electrostrictive_1996}
\nocite{rumble_density_2020}
\nocite{malitson_interspecimen_1965}
\nocite{rumble_speed_2020}

\printbibliography

\end{document}



\baselineskip24pt


\maketitle

\tableofcontents
\newpage

\renewcommand{\thefigure}{\arabic{figure}}
\renewcommand{\figurename}{Supplementary Fig.}
\renewcommand{\thetable}{S\arabic{table}}
\setcounter{figure}{0}
\renewcommand{\thesection}{S\arabic{section}}
\FloatBarrier
\section{Temporal calibration}\label{sec: SM_temporal_calibration}
The temporal calibration ensures that only one data-control pulse pair interacts in the optical fiber.
The isolation of a single stimulated Brillouin scattering (SBS) process allows us to rule out
any parasitic interactions, e.g., $a_\mathrm{D, 1}$ interacts with $a_\mathrm{C, 2}$.
In addition, the acoustic wave $b_1$ exists only at a single location in the optical fiber.
Note that for future applications of OREO, it might not be necessary to isolate
the interactions that strictly.

The selected PCF has a length of $40\,\mathrm{cm}$, corresponding to a pulse
travel time of about $2\,\mathrm{ns}$. Hence, by separating two pulses with
a deadtime of $2.5\,\mathrm{ns}$, one ensures single pulse interactions.
Furthermore, the path lengths of the data and control branch differ
in our setup due to the different devices in the corresponding path. Consequently,
we compensate for the different path lengths by electrically delaying the control pulses
with the arbitrary waveform generator (AWG) (see Figure~4 of the main text).

\begin{figure}[h!]
\centering
  \includegraphics[width=\textwidth]{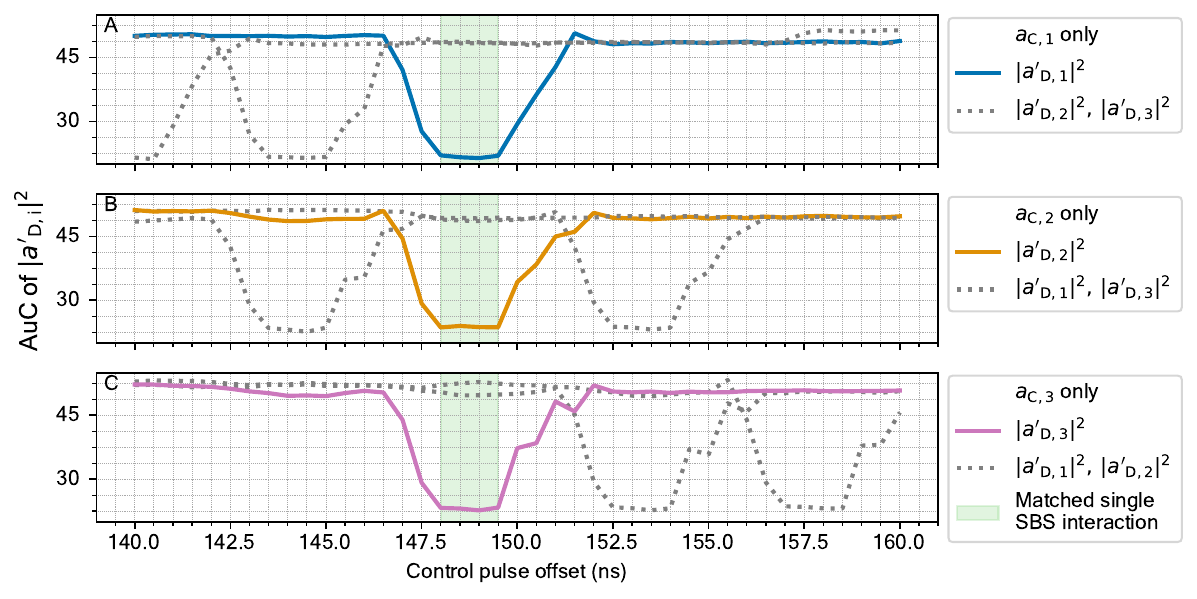}
    \caption{Results of OREO's temporal calibration to ensure that only a single and a matched data-control pulse pair interacts in the optical fiber. \textbf{A}, \textbf{B}, and \textbf{C} study the interaction of $a_\mathrm{C, 1}$, $a_\mathrm{C, 2}$ and $a_\mathrm{C, 3}$ with the different data pulses, respectively. In each plot, the colored solid line marks the desired interaction, e.g., in \textbf{A} it is $a_\mathrm{D, 1} \leftrightarrow a_\mathrm{C, 1}$. Moreover, the gray dashed lines represent unwanted interactions, e.g. in \textbf{A} those are $a_\mathrm{D, 2} \leftrightarrow a_\mathrm{C, 1}$ and $a_\mathrm{D, 3} \leftrightarrow a_\mathrm{C, 1}$.
    Possible offset values for the control pulses, where OREO performs only the desired isolated and matched interactions $a_\mathrm{D, i} \leftrightarrow a_\mathrm{C, i}$ are covered by the green rectangular in each plot.}
\label{fig: offset_calibration}
\end{figure}

We sweep the temporal offset of the control pulses, while keeping the data offset constant. In order to extract the optimal control offset, we measure the depletion of
the data pulses $a_\mathrm{D, 1}$, $a_\mathrm{D, 2}$, and $a_\mathrm{D, 3}$ for the case, where only one of the
control pulses $a_\mathrm{C, i}$ is active. This allows us to isolate the interaction
between the different pairs. The blue solid curve in Figure~\ref{fig: offset_calibration}~\textbf{A} represents the desired interaction $a_\mathrm{D, 1} \leftrightarrow a_\mathrm{C, 1}$. The dashed grey lines in the same plot represent the
interaction of $a_\mathrm{C, 1}$ either with $a_\mathrm{D, 2}$ or with $a_\mathrm{D, 3}$. As discussed, those
interactions are not of interest for the demonstration of OREO. Figure~\ref{fig:
offset_calibration}~\textbf{B} and \textbf{C} show the results, where $a_\mathrm{C, 2}$ and
$a_\mathrm{C, 3}$ only are active, respectively. The colored curve in both Figures shows the
behavior of the corresponding data pulse $a_\mathrm{D, 2}$ and $a_\mathrm{D, 3}$.
A green square in each plot marks the region, where we observe the desired interaction $a_\mathrm{D, i} \leftrightarrow a_\mathrm{C, i}$ and no parasitic interaction.
This region covers a time span of about $2\,\mathrm{ns}$ translating to the length of the fiber.
Consequently outside this zone, the data pulse $a_\mathrm{D, i}$ or the control pulse $a_\mathrm{C, i}$ takes only partially part in the SBS-process reducing its efficiency.
Based on the results presented in Figure~\ref{fig: offset_calibration}, we chose $149\,\mathrm{ns}$ as offset for the control pulses.
\FloatBarrier
%
%
%
\FloatBarrier
\section{Intrinsic correlation check}
The all-optical control is one of OREO's unique features. In order to validate that
the connection between the different data pulses is induced by the SBS-process alone, we check the
correlation of the different data pulses. Therefore, we sweep the amplitude of
data pulses $a_\mathrm{D, 1}$ and $a_\mathrm{D, 2}$, while checking the impact on all subsequent data pulses.
We perform this study twice with different settings. In the first sweep, the
the control pulses are turned off and in a second sweep, the control pulses are on but temporally detunded by $351 \, \mathrm{ns}$ to the optimal offset.

\begin{figure}[t!]
\centering
  \includegraphics[width=.85\textwidth]{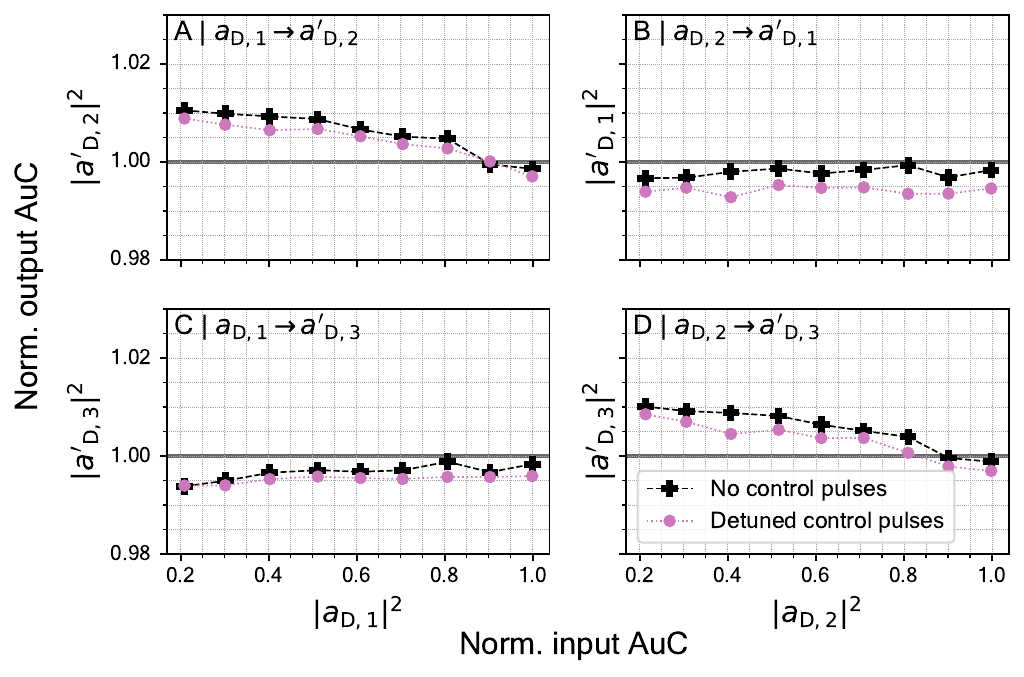}
    \caption{Results for the intrinsic correlation check of OREO. In \textbf{A} and \textbf{C} we can see a small correlation between the input amplitude of data pulse $a_\mathrm{D, 1}$ and the output amplitudes $a'_\mathrm{D, 2}$ and $a'_\mathrm{D, 3}$, respectively. The reason for this dynamic could be the RF-amplifier located between the AWG and the intensity modulator. A similar dynamic between subsequent pulses occurs if we change the data pulse $a_\mathrm{D, 2}$ and check its impact on the output $a'_\mathrm{D, 3}$ (see~\textbf{D}). Again the output amplitude increases for lower inputs. As expected, we can see no connection between the input of $a_\mathrm{D, 2}$ and the output $a'_\mathrm{D, 1}$.
    However, we see that the EDFA, although detuned, reduces for all four cases the amplitude of the output data pulses. This effect might result from the EDFA's amplified spontaneous emission (ASE).
    Overall, the intrinsic correlation is significantly lower as the one induced via ORE and can, therefore, be neglected.}
  \label{fig: result_correlation_check}
\end{figure}

If we sweep the amplitude of data pulse $a_\mathrm{D, 1}$, we can observe a small impact on the output pulses $a'_\mathrm{D, 2}$ and $a'_\mathrm{D, 3}$ as shown by Figure~\ref{fig: result_correlation_check}~\textbf{A} and \textbf{C}, respectively.
The increase of the output amplitude $a'_\mathrm{D, 2}$ for lower values of $a_\mathrm{D, 1}$ (see Fig.~\ref{fig: result_correlation_check}~\textbf{A}) could be caused by the RF-amplifier which is placed between the
arbitrary waveform generator (AWG) and the intensity modulator, which is used for the pulse generation. If the electrical amplitude of the data pulse $a_\mathrm{D, 1}$ decreases the RF-amplifier has more energy to feed into the
subsequent data pulse $a_\mathrm{D, 2}$. This could also explain, why the amplitude of $a'_\mathrm{D, 3}$ decreases in the same time
(see Fig.~\ref{fig: result_correlation_check}~\textbf{C}). The same effect can most probably be observed for
the sweep  data pulse $a_\mathrm{D, 2}$. Here, the output amplitude of the subsequent data pulse $a'_\mathrm{D, 3}$ also increases
for lower input values of $a_\mathrm{D, 2}$ (see Fig.~\ref{fig: result_correlation_check}~\textbf{D}).
In addition, we do not observe any effect of $a_\mathrm{D, 2}$ on the output $a'_\mathrm{D, 1}$ as illustrated in
Figure~\ref{fig: result_correlation_check}~\textbf{B},
indeed, the data pulse remains almost constant. An overall drift of the system could explain, why the AuC value is slightly below $1$. Moreover, if one compares the two measurements in Figure~\ref{fig: result_correlation_check}~\textbf{B}, one notices that the detuned control signal slightly depletes the data signal. Most probably this is could be an effect of the EDFA's amplified spontaneous emission (ASE).

In comparison to the results discussed in the main part, we can conclude that the measured intrinsic correlation are neglectable. However, one should keep in mind the data pulse depletion induced by the ASE of the EDFA, which might becomes stronger for higher pump powers of the EDFA.
\FloatBarrier

%
%
%
\section{Performance evaluation}
This section asses the performance of OREO in terms of its speed and energy consumption.
The key parameters of OREOs performance are the deadtime $dt$ between
two pulse pairs, and the pulse width $\tau$.The lower boundary of the deadtime can be estimated as $dt > \frac{L\, n_{\mathrm{eff}}}{\mathrm{c}_{0}}$,
where $L$ is the waveguide's length, $n_{\mathrm{eff}}$ is it effective refractive index, and
$\mathrm{c}_{0}$ is the speed of light. In total, the overall number of \color{black}operations (OPS) \color{black} which can be
achieved in one second is:
\begin{equation}
  \label{eq: performance_OREO_single_frquency}
  C_{\mathrm{SF}} = \frac{1}{dt + \tau}.
\end{equation}
For a short fiber, one could operate OREO with a deadtime of $dt=500\,\mathrm{ps}$ and a pulse width of $\tau = 250 \,\mathrm{ps}$, yielding
$C_{\mathrm{SF}}= 1.33\,\mathrm{GROPS}\,\mathrm{s}^{-1}$.
The frequency selectiveness of Brillouin scattering allows OREO to process $M$ frequencies simultaneously, extending the performance of OREO further to $C_{\mathrm{MF}}=M\cdot C_{\mathrm{SF}}$.
The number of frequencies that can be utilized by the scheme depends on the spectral width of the pulses.
For example, a channel separation of $3.6\,\mathrm{GHz}$ and a photodetector with a bandwidth of
 $100\,\mathrm{GHz}$ boost the performance of the discussed example by a factor of $28$ to $C_{\mathrm{MF}}\approx 28 \cdot C_{\mathrm{SF}}=36.4\,\mathrm{GROPS}\,\mathrm{s}^{-1}$.

The current approach of OREO requires about $P_{\mathrm{pulse}}=126\,\mathrm{mW}$ of optical power
per control pulse $a_\mathrm{C, i}$ in the PCF, translating to a pulse energy of $E_{\mathrm{pulse}}=126\,\mathrm{nJ}$
for the applied rep. rate of $1\,\mathrm{MHz}$.
Hence, OREO has an energy efficiency of about $ \eta\approx 8\,\mathrm{MROPS}\,{\mathrm{J}}^{-1} $.

A control pulse recycling scheme can enhance OREO's energy efficiency significantly.
Such a scheme can be implemented because a control pulse is not depleted through
the SBS process, but amplified. Hence, it offers the possibility to re-use it for subsequent computational steps.  
In addition, the experimental scheme presented in Figure~4 of the main text offers
also an output for the control pulse.
In the use case discussed above, a $10\,\mathrm{cm}$ delay line between the
control input and output of OREO would already be enough for implementing a
recycling scheme. Under the assumption of an low loss recycling scheme, where the loss
is compensated through the control pulse amplification of the SBS process, the control pulse recycling could increase the efficiency by nine
orders of magnitude:
\begin{align}
\begin{aligned}
  \label{eq: efficiency_recycling}
  \eta_{\mathrm{recy}}&=\frac{1}{(dt +\tau) \, E_{\mathrm{pulse}}} = 2.3 \, \frac{\mathrm{PROPS}}{\mathrm{J}} \approx 10^9\cdot\eta, \\
  \mathrm{using} &\quad dt =2.5\,\mathrm{ns}, \, \tau =1\,\mathrm{ns}
  \end{aligned}
\end{align}
For OREO with a long short-term memory that employs a deadtime of $dt=500\,\mathrm{ps}$
and a pulse width of  $\tau=250\,\mathrm{ps}$ the computational efficiency can be increased to $\eta_{\mathrm{recy}} =10.6 \,\mathrm{PROPS}\,{\mathrm{J}}^{-1}$.
Note that a waveguide with a high optoacoustic gain could further increase the efficiency of OREO.
%
%
%
%
%
%
\FloatBarrier
\section{All-optical control}
The control pulses $a_\mathrm{C, i}$ can be used to tune the strength of the
optoacoustic interaction. Known from an optoacoustic memory~\autocite{zhu_stored_2007, merklein_chip-integrated_2017}, the power of the control
pulses dictates the degree of depletion and, therefore, the relative amount of acoustic
that is created. Hence, the control pulses can be used to optimize the interaction
for the desired need or to modify the light-sound interaction on pulse-by-pulse level.

\begin{figure}[h!]
\centering
  \includegraphics[width=\textwidth]{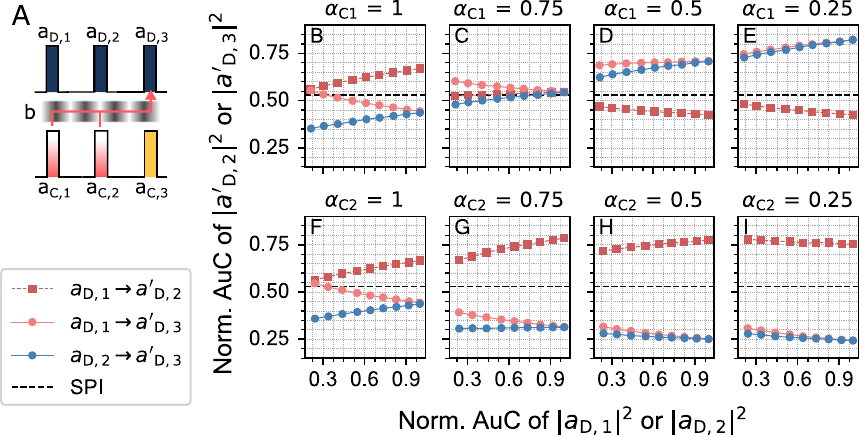}
    \caption{Study of OREO's capability to control the recurrent operation completely optically on pulse-by-pulse basis. The pulse pairs are seperated by a  deadtime of $dt=2.5\,\mathrm{ns}$.
            \textbf{A} - Schematic illustration of the control pulse sweep that investigates the impact of different control pulse amplitudes
             on the recurrent interaction. \textbf{B} to \textbf{E} - Experimental results of reducing the amplitude of control pulse $a_\mathrm{C, 1}$.
             The relative change in control amplitude is given by $\alpha_{C1}$. This measurement isolates the impact of the acoustic wave $b_1$ on subsequent SBS processes. \textbf{F} to \textbf{I} - Experimental results of reducing the amplitude of control pulse $a_\mathrm{C, 2}$. This measurement isolates the impact of the acoustic wave $b_2$.}
\label{fig: effect_amplitude_control_pulse}
\end{figure}

In the following, we study the all-optical control with three data-control pairs (deadtime
$2.5\,\mathrm{ns}$), where we vary the amplitude either from control pulse $a_\mathrm{C, 1}$ or $a_\mathrm{C, 2}$,
while keeping the remaining control pulse constant (see Fig.~\ref{fig: effect_amplitude_control_pulse} \textbf{A}). The results for four
amplitudes scales, namely, full ($\alpha_{Ci}=1$), three quarters ($\alpha_{Ci}=0.75$), half ($\alpha_{Ci}=0.5$),
and quarter ($\alpha_{Ci}=0.25$) are shown in Figure~\ref{fig: effect_amplitude_control_pulse}. 
As the acoustic control amplitude dictates the degree of depletion of the data pulse, we can use it
to isolate the impact of certain acoustic contributions.
For example, the amplitude sweep of control pulse $a_\mathrm{C, 1}$ (see Fig.~\ref{fig: effect_amplitude_control_pulse} \textbf{B} to \textbf{E})
displays the impact of the acoustic interference when one studies the interaction $a_\mathrm{D, 1} \to a'_\mathrm{D, 2}$.
In the process of reducing the control pulse amplitude, one can see the transition from an inhibition to annihilation to amplification of the
interaction $a_\mathrm{D, 2} \leftrightarrow a_\mathrm{C, 2}$ in the plots \textbf{B}, \textbf{C}, and \textbf{D}, respectively. An equivalent dynamic can be observed for
the acoustic link $a_\mathrm{D, 1} \to a'_\mathrm{D, 3}$, which is influenced by the acoustic interference of $b_1$ and $b_2$.
Here, the acoustic link first amplifies the interaction of $a_\mathrm{D, 3} \leftrightarrow a_\mathrm{C, 3}$, then annihilates it and, finally, inhabits it as one can
see in the plots \textbf{B}, \textbf{D}, and \textbf{E}, respectively. Next, the acoustic wave $b_1$ serves the
acoustic link $a_\mathrm{D, 2} \to a'_\mathrm{D, 3}$ as an amplifier because the degree of depletion of $a_\mathrm{D, 3}$ shrinks for higher control pulse attenuations $\alpha_{C1}$ (see Fig.~\ref{fig: effect_amplitude_control_pulse} \textbf{B} to \textbf{E}).
Next, the amplitude sweep of control pulse $a_\mathrm{C, 2}$ (see Fig.~\ref{fig: effect_amplitude_control_pulse} \textbf{F} to \textbf{I})
reveals information about the influence of the acoustic wave $b_2$. This interaction
inhibits the acoustic link between $a_\mathrm{D, 1}\to a'_\mathrm{D, 3}$ because the depletion of $a'_\mathrm{D, 3}$ increases with lower amplitudes of $a_\mathrm{C, 2}$ (see Fig.~\ref{fig: effect_amplitude_control_pulse} \textbf{F}).
In addition, the acoustic link $a_\mathrm{D, 2}\to a'_\mathrm{D, 3}$ shows also the effect of the acoustic interference between $b_1$ and $b_2$: for
$\alpha_{C1}=1$, the SBS-process is inhibited; for $\alpha_{C1}=0.75$ it is annihilated; and for $\alpha_{C1}=0.5$ it is amplified (see Fig.~\ref{fig: effect_amplitude_control_pulse} \textbf{F} to \textbf{H}, respectively).

Overall, the control pulses $a_\mathrm{C, i}$ serve OREO as an additional degree of freedom to manipulate the recurrent operation on pulse-by-pulse basis.
\FloatBarrier
%
%
%
%
%
%
\FloatBarrier
\section{OREO with a highly nonlinear fiber}
\begin{figure}[h!]
\centering
  \includegraphics[width=\textwidth]{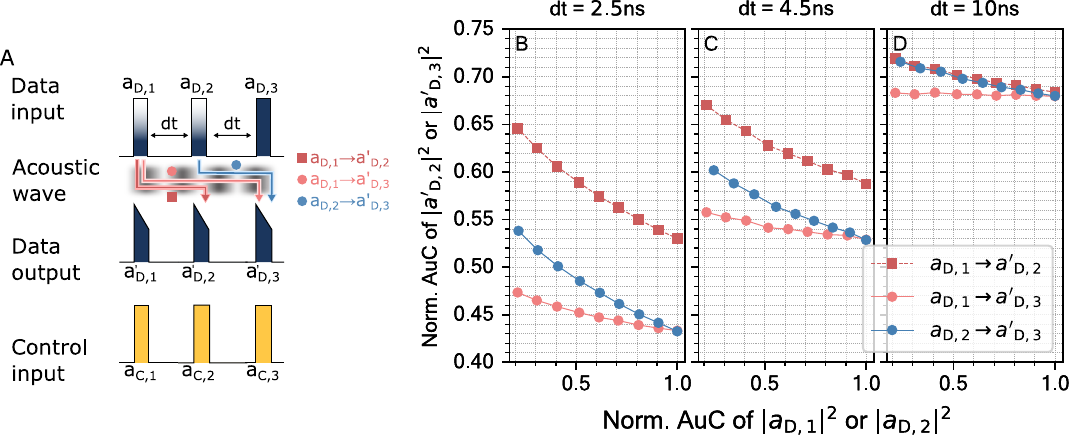}
\caption{Observing OREO's optoacoustic linking and pulse skipping capabilities in a HNLF. \textbf{A} - Schematic illustration of the amplitude sweep that investigates how different optical states are passed between the optical data pulses $a_\mathrm{D, i}$ via an acoustic wave $b$.
\textbf{B} to \textbf{D} - Experimental results of the amplitude sweep in a HNLF. While $a_\mathrm{D, 1}$ and $a_\mathrm{D, 2}$ are changed, their impact on the subsequent pulses $a_\mathrm{D, 2}$ and $a_\mathrm{D, 3}$ are studied for different deadtimes $dt$. Each interaction $a_\mathrm{D, i} \leftrightarrow a_\mathrm{C, i}$ creates an acoustic wave that interacts with pre-exisiting ones, eventually, effecting following SBS processes.} \label{fig: HNLF_amp_sweep}
\end{figure}

In the following, we demonstrate OREO in a $\approx 43\,\mathrm{cm}$ long
highly nonlinear fiber (HNLF). Analogous to the PCF measurement, we launch three consecutive data-control pulse pairs into the fiber
and vary the amplitude of $a_\mathrm{D, 1}$ and $a_\mathrm{D, 2}$ (see Figure~\ref{fig: HNLF_amp_sweep} \textbf{A}). In contrast to the PCF study, we investigate OREO's dynamic without a frequency detuning, i.e., the frequency difference between data and control pulses matches the Brillouin frequency. Hence, we drop the influence of the acoustic phase. We study the dynamic again for different deadtimes. 
The recurrent dynamic is less complex as the one of the PCF due to the lack of acoustic interference. As we can see from Figure~\ref{fig: HNLF_amp_sweep} \textbf{B}, each SBS process is enhanced by the previous ones. 
As a result the output amplitude of $a'_\mathrm{D, 2}$ is higher as the one of $a'_\mathrm{D, 3}$ for all interactions. In contrast to the PCF, this dynamic does not change for different deadtimes (see from Figure~\ref{fig: HNLF_amp_sweep} \textbf{C}). The overall higher output amplitudes $a'\mathrm{D,i}$ 
for the different acoustic links ($a_\mathrm{D, i} \to a'\mathrm{D, j}$) can be explained
with the decay of the acoustic wave. Although the link $a_\mathrm{D, 1} \to a'_\mathrm{D, 3}$
touches the acoustic lifetime limit, we are still able to observe a recurrent interaction. This link ($a_\mathrm{D, 1} \to a'_\mathrm{D, 3}$) vanishes clearly as soon as we separate the pulse pairs by the acoustic lifetime (see from Figure~\ref{fig: HNLF_amp_sweep} \textbf{D}).
In this case, we can only observe nearest neighbour interactions $a_\mathrm{D, i} \to a'_\mathrm{D, i+1}$.  
OREO achieves in our HNLF a maximum dynamic range of $\approx 20\,\%$. A numerical study of the HNLF behavior is depicted in Figure~\ref{fig: numerical_study_abc_OREO_HNLF}
\FloatBarrier

%
%
\section{Theory of OREO}
OREO employs stimulated Brillouin scattering (SBS),
which is well-studied in literature~\autocite{wolff_brillouin_2021, boyd_nonlinear_2008, agrawal_nonlinear_2013, eggleton_brillouin_2022, eggleton_brillouin_2022-1}.
In the following, we use those references to elaborate on the theoretical foundation of OREO. Note that we restrict ourself
to backward SBS.

\begin{figure}[h!]
\centering
  \includegraphics[width=.55\textwidth]{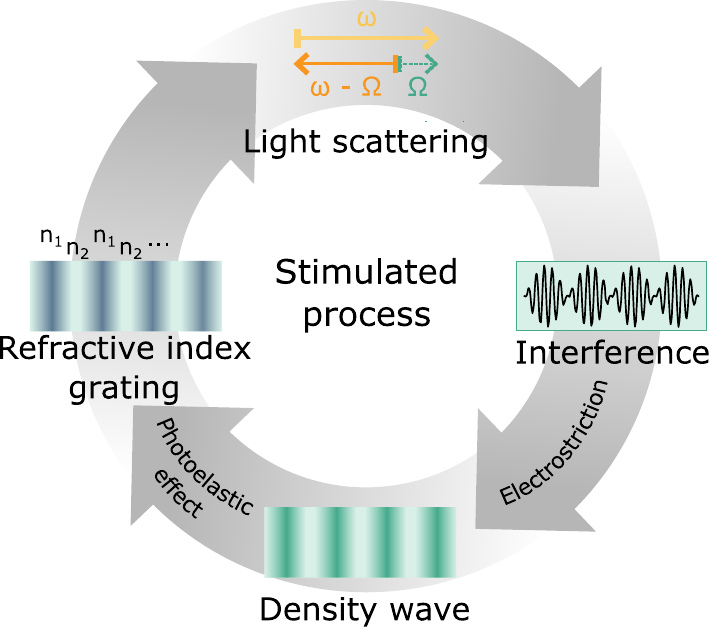}
    {\caption{Schematic illustration of the feedback loop between the acoustic and optical waves. An initial light wave scatters from acoustic phonons and is shifted by the Brillouin frequency $\Omega$.
    The initial and backscatter light field interfere with each other and create a moving interference pattern. Electrosriction transforms this interference pattern to a moving density wave inside the waveguide, which changes the optical properties of the medium due to the photoelastic effect. The resulting density wave enhances the scattering efficiency and, thus, enables a stimulated process.}
\label{fig: illustration_SBS_feedback}}
\end{figure}

SBS is a third-order nonlinear effect that  describes the coherent interaction of light
with sound. This interaction is established by two effects, namely, electrostriction and
the photoelastic effect. Electrostriction labels the creation of dipole moments inside a
medium by a light field. The dipole moments experience an attractive force, pulling them
towards the electromagnetic field and changing the medium's density. Eventually, the
density changes form an acoustic wave, i.e., density wave, that changes the dielectric
properties of the medium due to the photoelastic effect.
The optical field experiences the periodic nature of the acoustic wave as a grating, from which it inelastically scatters.
The scattered light is red or blue shifted by the Brillouin frequency $\Omega$ of the acoustic field and depends on the propagating direction of the acoustic wave.
The Brillouin frequency $\Omega$ lays in the range of several gigahertz.
Furthermore, the initial and the backscattered light field interfere with each other and, thereby,
accelerate the forming of the acoustic wave, which in turn increases the amount of light
that is backscattered (see Fig.~\ref{fig: illustration_SBS_feedback}). Hence, this feedback loop forms a stimulated process that
significantly enhances the efficiency of Brillouin scattering. Indeed, stimulated Brillouin
scattering is several orders of magnitude stronger then the Raman- or the Kerr-effect~\autocite{wolff_brillouin_2021}.

According to reference~\autocite{wolff_brillouin_2021}, the dynamics of the SBS process can be described with the
one-dimensional optical and acoustic wave equation, which are noted in equations~\eqref{eq: optical_ave_equation} and \eqref{eq: acoustic_wave_equation}, respectively.
\begin{align}
          \frac{\partial^2}{\partial z^2}E - \frac{n^2}{c_0^2}\frac{\partial^2}{\partial t^2}E &= \frac{4\pi}{c_0^2} \frac{\partial^2}{\partial t^2}P, \label{eq: optical_ave_equation} \\
          \frac{\partial^2}{\partial t^2}b-\Gamma'\frac{\partial^2}{\partial z^2}\frac{\partial}{\partial t}b - v_\mathrm{ac} \frac{\partial^2}{\partial z^2}b &= \nabla \vec{f} \label{eq: acoustic_wave_equation}.
\end{align}
The set of equations~\eqref{eq: optical_ave_equation} and \eqref{eq: acoustic_wave_equation} use the one dimensional electric field $E$, the refractive index $n$, the speed of light $c_0$, the polarization $P$, the acoustic damping term $\Gamma'$ and the acoustic velocity $v_\mathrm{ac}$. The polarization $P$ and the driving term $\vec{f}$ of the acoustic field can connected to electrostiction, giving equations \eqref{eq: optical_drive_term} and \eqref{eq: acoustic_drive_term}, respectively.
\begin{align}
          P &= \frac{1}{4\pi \rho_0} b\,E, \label{eq: optical_drive_term} \\
          \vec{f} &= -\gamma_\mathrm{e} \nabla \frac{\left\langle\vec{E}\cdot \vec{E}\right\rangle}{8\pi} \label{eq: acoustic_drive_term}.
\end{align}
With the density of the waveguide $b_0$ and the electrostrictive constant $\gamma_\mathrm{e}$.

Applying a plain wave ansatz for both the optical and acoustic wave, gives a set of nonlinear  partial
differential equations. Those cannot be solved analytically without applying several approximations.
According to reference~\autocite{wolff_brillouin_2021}, one get a set of linear coupled partial differential
equations by using the rotating frame and slowly-varying wave approximation, yielding the known coupled mode
equations of the SBS process, which include two optical equations for the data
wave $a_\mathrm{D}$ and the control wave $a_\mathrm{C}$ and one equation for the acoustic
field $b$:
\begin{align}
\label{eq: OREO_CME}
\begin{aligned}
      \left(\frac{n_\mathrm{eff}}{c_{0}} \frac{\partial}{\partial \, t} + \frac{\partial}{\partial \, z}\right) a_{\mathrm{D}} &= - \mathrm{i} \frac{\gamma_\mathrm{e} \left(\omega_\mathrm{C} + \Omega\right)^2 }{2 \omega_\mathrm{D} \rho_0 c_0 n_\mathrm{eff}} a_\mathrm{C} b \, \mathrm{e}^{\mathrm{i}\left[\left(\omega_\mathrm{C} + \Omega\right) t - \left(q-k_\mathrm{C}\right) z \right]} \\
      \left(\frac{n_\mathrm{eff}}{c_{0}} \frac{\partial}{\partial \, t} - \frac{\partial}{\partial \, z}\right) a_{\mathrm{C}} &= - \mathrm{i} \frac{\gamma_\mathrm{e}  \left(\omega_\mathrm{D} - \Omega\right)^2 }{2 \omega_\mathrm{C} \rho_0 c_0 n_\mathrm{eff}} a_\mathrm{D} b^* \, \mathrm{e}^{\mathrm{i}\left[\left(\omega_\mathrm{D} - \Omega\right) t - \left(k_\mathrm{D} - q\right) z \right]}, \\
      \left(\frac{\partial}{\partial \, t} + v_\mathrm{ac} \frac{\partial}{\partial \, z} + \Gamma_\mathrm{b} \right) b &= - \mathrm{i}\frac{ \gamma_\mathrm{e} \left(k_\mathrm{D} + k_\mathrm{C}\right)^2 }{8 \pi q v_\mathrm{ac}} a_\mathrm{D} a_\mathrm{C}^* \,\mathrm{e}^{\mathrm{i}\left[\left(\omega_\mathrm{D} - \omega_\mathrm{C}\right) t - \left(k_\mathrm{D}+k_\mathrm{C}\right) z \right]}.
\end{aligned}
\end{align}
With the effective refractive index of the waveguide $n_\mathrm{eff}$, the speed of light $c_0$, the elecostrictive constant $\gamma_\mathrm{e}$, the density of the waveguide $b_0$, the acoustic group velocity in the waveguide $v_\mathrm{ac}$, the acoustic linewidth $\Gamma_\mathrm{b}$, the optical wave vectors $k_\mathrm{c, d}$, the acoustic wavevector $q$ and the frequency relation between the fields $\omega_\mathrm{d} = \omega_\mathrm{c} + \Omega+\Delta\omega$. Note that this set of equations capture also the acoustic phase introduced by an optical detuning from the Brillouin frequency: $\Delta\omega\neq 0$.
The coupled mode equations can also be derived from the Hamiltonian given in equation (1) of the main text by applying the Heisenberg equation~\autocite{zhang_quantum_2023}.


Next, we numerically study the dynamics of the coupled mode equations (CME) \eqref{eq: OREO_CME} with the technique presented in
Reference~\autocite{de_sterke_nonlinear_1991}. This method uses the
characteristics of the data and control wave
$\frac{n_\mathrm{eff}}{c_0} \pm z = 0$ and solve the CME along the
$\frac{c_0 }{n_\mathrm{eff}}t \pm z$-direction via an implicit
Runge-Kutta scheme. As this approach only tackles the optical part
of the CME, we also employ the Euler
method~\autocite{butcher_numerical_2005} to solve the acoustic part of the CME.
Therefore, we drop the space-derivative of the acoustic
mode, assuming a stationary acoustic field.
In order to run the simulation we feed the parameters shown in Table~S1 into the numerical framework.
\begin{table}
  \centering
  \caption{Simulation parameters for the spectral-analysis of SBS.}
  \label{tab: sim_params_spectral_analysis_SBS}
  \begin{tabular}{l c c c}
    \hline
    Name & Symbol & Value & Reference\\
    \hline
    Elecostrictive const. & $\gamma_\mathrm{e}$ & $1\, \mathrm{m} \,\mathrm{W}^{-1}$ &
    \autocite{buckland_electrostrictive_1996} \\
    Density & $\rho_0$ & $2.2\cdot 10^3 \, \mathrm{g}\,\mathrm{m}^{-3} $ & \autocite{rumble_density_2020} \\
    Refractive index & $n$ & 1.44 & \autocite{malitson_interspecimen_1965} \\
    Acoustic group velocity & $v_\mathrm{ac} $ & $6.3\cdot 10^3 \mathrm{m}\,\mathrm{s}^{-1}$ & \autocite{rumble_speed_2020}\\
    Effective area & $A_\mathrm{eff}$  & $1.5\cdot 10^{-12} \mathrm{m}^{-2}$  & -\\
    Acoustic lifetime & $\tau \propto \Gamma_\mathrm{B}^{-1}$  & $8.8 \, \mathrm{ns}$  & -\\
    \hline
\end{tabular}
\end{table}

Figure~\ref{fig: numerical_study_abc_OREO_HNLF} depicts the simulation results
of OREO in a highly nonlinear fiber.
Although, we observe an overall agreement with the measurements results of the
HNLF (see Fig. \ref{fig: HNLF_amp_sweep}), the simulation show a
linear behavior of the acoustic link for $2.5\,\mathrm{ns}$ (compare Fig.~\ref{fig: HNLF_amp_sweep} \textbf{B} and Fig.~\ref{fig: numerical_study_abc_OREO_HNLF} \textbf{A}).
\begin{figure}[h!]
\centering
  \includegraphics[width=.75\textwidth]{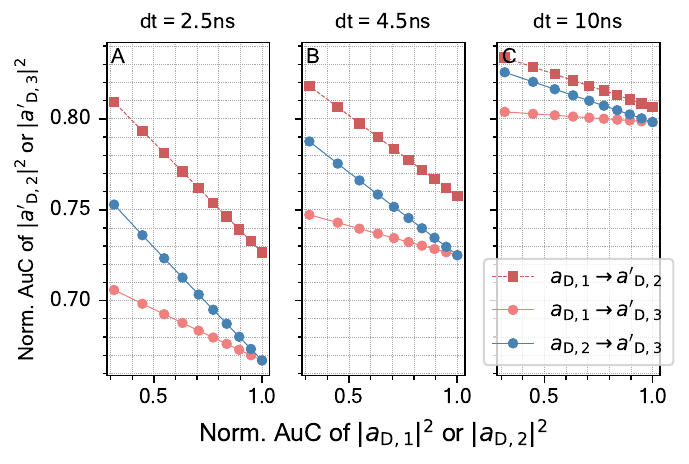}
\caption{Numerical results of OREOs dynamic for a highly nonlinear fiber. The simulation results show agree almost completely with the measurements shown in Figure~\ref{fig: HNLF_amp_sweep}.}
\label{fig: simulation_HNLF_behavior}
\end{figure}
%
%
\section{Numerical study of the $abc$-pattern prediction}
In the following, we apply the simulation framework to study OREO's $abc$-pattern recognition performance in a highly nonlinear fiber (HNLF).
In comparison to the $abc$-measurements, we operate OREO in-resonance ($\omega_\mathrm{d}=\omega_\mathrm{c}+\Omega+\Delta\omega$, $\Delta\omega = 0$) in order to evaluate OREO's fundamental response.
We study the $abc$-pattern recognition task for different pulse lengths ($pw$), deadtimes ($dt$), acoustic lifetimes $\tau$, and experimental
precisions. The first three variables are input parameters of the simulation framework, whereas the latter one has to be considered
in post-processing. Therefore, we simulate OREO's dynamic for each of the $27$ patterns considering a certain ($pw$,
$dt$, $\tau$)-setting. This yields $27$ AuC-values for the Eval'-pulse, which serve as mean value used to draw $n=1000$
samples from a Gaussian distribution. The width of the Gaussian distribution is determined by the mean standard deviation
acquired in the $abc$-pattern recognition experiment: $\overline{\sigma}_\mathrm{exp} = 0.13 \, \pm \, 0.01 \, \mathrm{abr. u.}$.
We numerically increase the precision of our experiment by decreasing
$\overline{\sigma}_\mathrm{exp}$, e.g., doubling the precision corresponds to
$\overline{\sigma}_\mathrm{exp} /2$. In total, we generate a dataset of size
$27000$, which we feed into a \textsc{RandomForst} (RFC) in the same way as for the experiment, which then returns a predictive accuracy.

\begin{figure}[h!]
\centering
  \includegraphics[width=\textwidth]{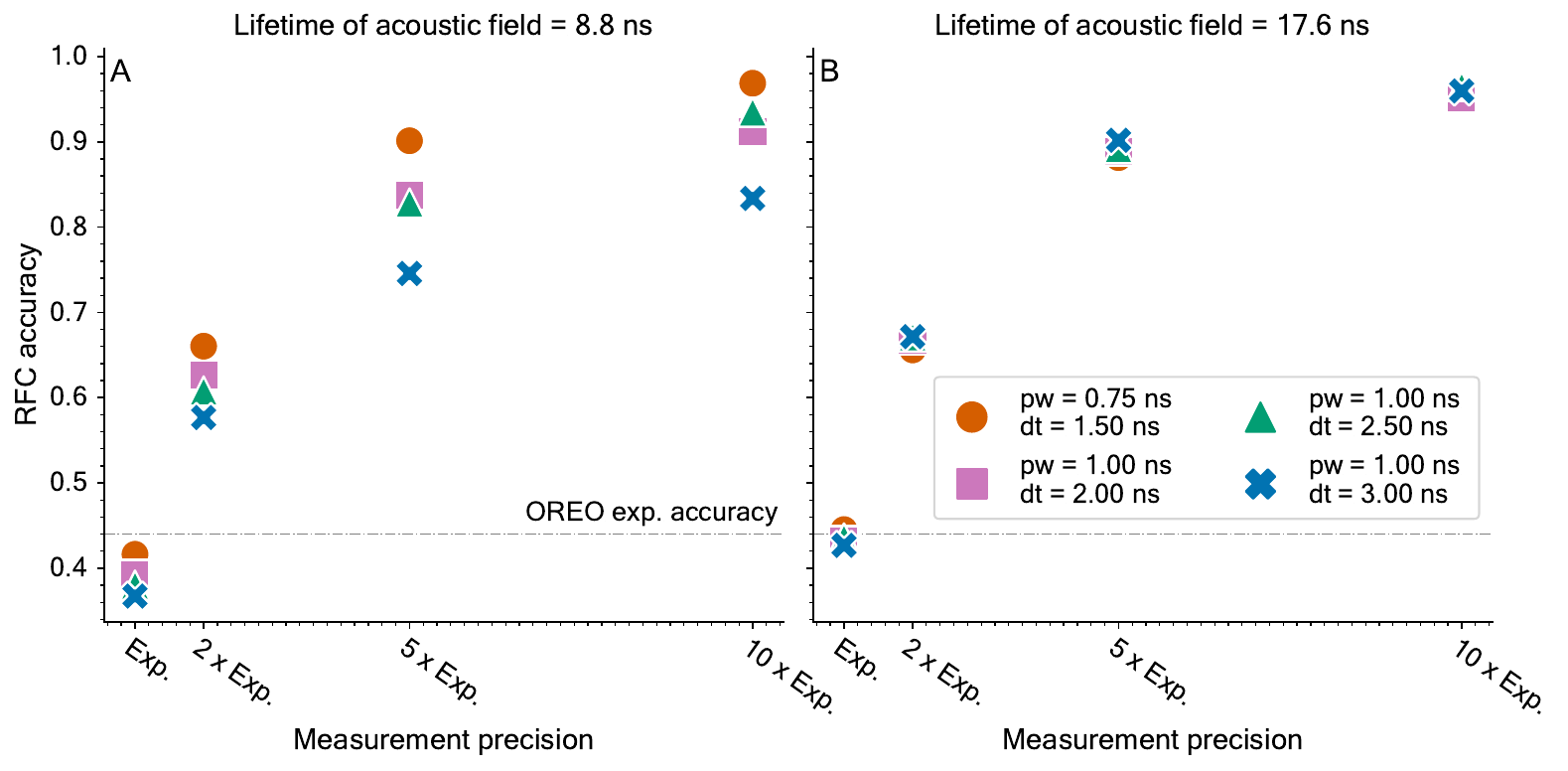}
 {\caption{Numerical study of the $abc$-pattern prediction task. We generate a dataset by simulating OREO's behavior for the different $27$-patterns, which
 gives an estimate for Eval'. Combined with the precision obtained in the experiment, we draw for each pattern $1000$ times from a Gaussian
 distribution, where the estimate for Eval' represents its mean value. We perform
 this procedure for different pulse widths ($pw$), deadtimes ($dt$), and acoustic
 lifetimes. We simulate different experimental precision by reducing the value
 extracted from the experimental distributions, e.g., doubling the precision would
correspond to $\overline{\sigma}_\mathrm{exp} /2$. In the most optimized case OREO achieves together with the RFC an classification accuracy of $97\,\%$.}
\label{fig: numerical_study_abc_OREO_HNLF}}
\end{figure}

Figure~\ref{fig: numerical_study_abc_OREO_HNLF} shows the results of the numerical
study for different pulse widths, deadtimes, and experimental precision at each of two acoustic lifetimes. The results suggest that the experimental precision has the highest impact on the RFC's accuracy.
A five time increase in precision compares to the measurement results of the $ab$-measurement, which we discussed in the main text.
In the most optimized case, the RFC achieves an accuracy of $92\,\%$ for the pulse width and deadtime used in the experiment.
In addition, in the case of the standard acoustic lifetime of $8.8\,\mathrm{ns}$ as measured in experiment (Fig.~\ref{fig:
numerical_study_abc_OREO_HNLF} \textbf{A}), we observe that a shorter
deadtime $dt$ is beneficial for the performance of OREO. However, the influence of the investigated deadtime span decreases for the case of increased acoustic lifetime due to the slower acoustic decay. This explains why the different markers
are less distributed in Figure~\ref{fig: numerical_study_abc_OREO_HNLF} \textbf{B}. In addition, it suggests that a high acoustic lifetime could allow OREO to capture even longer pattern, i.e., more complex context carried by the optical domain. In this case, the precision of OREO can be pushed to $97\,\%$

%
%
\clearpage
\section{References}
\printbibliography[heading=none]